\documentclass[a4paper,aps,pre,twocolumn,groupedaddress,showkeys,floats,floatats,floatfix]{revtex4}
\usepackage[latin1]{inputenc}
\usepackage{graphicx}  
\usepackage{dcolumn}   
\usepackage{bm}        
\usepackage{natbib}
\usepackage{latexsym}
\usepackage{mathrsfs}
\usepackage{amssymb}
\usepackage{amsmath}
\usepackage{amscd}
\usepackage{color}
\usepackage{verbatim}
\begin{document}
\title{Soft wall effects on interacting particles in billiards} 
\author{H.~A.~Oliveira, C.~Manchein and M.~W.~Beims}
\email[E-mail address:~]{mbeims@fisica.ufpr.br}
\affiliation{Departamento de F\'\i sica, Universidade Federal do Paran\'a,
         81531-990 Curitiba, PR, Brazil}


\begin{abstract}
The effect of physically realizable wall potentials (soft walls) on the 
dynamics of two interacting particles in a one-dimensional (1D) billiard 
is examined numerically. The 1D walls are modelled by the error function and 
the transition from hard to soft walls can be analyzed continuously by 
varying the softness parameter $\sigma$. For $\sigma\rightarrow0$ the 
1D hard wall limit is obtained and the corresponding wall force on the 
particles is the $\delta$-function. In this limit the interacting 
particle dynamics agrees with previous results obtained for the 1D hard walls. 
We show that the two interacting particles in the 1D soft walls model is 
equivalent to one particle inside a {\it soft} right triangular billiard. Very 
small values of $\sigma$ substantiously change the dynamics inside the billiard 
and the mean finite-time Lyapunov exponent decreases significantly as the 
consequence of regular islands which appear due to 
the {\it low-energy double collisions} (simultaneous particle-particle-1D wall 
collisions). The rise of regular islands and sticky trajectories induced 
by the 1D wall softness is quantified by the  number of occurrences of the most 
probable finite-time Lyapunov exponent. On the other hand, chaotic motion in
the system appears due to the {\it high-energy double collisions}. In general 
we observe that the mean finite-time Lyapunov 
exponent decreases when $\sigma$ increases, but the number of occurrences 
of the most probable finite-time Lyapunov exponent increases, meaning that 
the phase-space dynamics tends to be more ergodic-like. Our results suggest 
that the transport efficiency of interacting particles and 
heat conduction in periodic structures modelled by billiards, will 
strongly be affected by the smoothness of physically realizable walls.
\end{abstract}



\maketitle

\section{Introduction}
\label{Introduction}

Although physically realizable potentials are inherently soft, most
billiard models used in the literature have hard walls. For example, the 
Sinai billiard~\cite{sinai}, the Bunimovich stadium~\cite{buni} or the 
Annular billiard~\cite{anular}, among others, have hard walls and were 
used successfully to study the fundamental properties of classical and 
quantum chaotic systems. In such 
models the chaotic motion of the single particle dynamics arises as the
consequence of the spatial billiard geometry. The question now is about the 
effect and importance of physically realizable potentials on the particles 
dynamics inside the billiards. Some works in this direction have shown that 
introducing soft walls do not destroy trajectories found in the 
hard-wall limit \cite{ana} and may induce the appearance of regular islands 
in phase space \cite{turaev98,turaev99,turaev03}. Such regular islands inside 
the chaotic sea induce a ``sticky''(or trapped) motion, which is a common 
phenomenum in conservative systems \cite{zas}. They arise from broken 
Kolmogorov-Arnold-Moser (KAM) \cite{lieberman} curves and generate a rich 
dynamics in quasi-integrable systems~\cite{zas}.  In the context of soft 
walls, the sticky motion has been observed theoretically and experimentally 
in the one particle atom-optic billiard \cite{kaplan1,kaplan2} and has 
shown to affect the quantum conductance in the soft wall microwave 
billiard~\cite{weingaertner}.

Interacting many-particles systems in soft wall billiards is the next 
step to be studied. Such billiards are interesting not only from the
fundamental point of view in nonlinear systems~\cite{donnay} but also in 
many applications. Two recent examples can be mentioned:  The effect of 
wall roughness in granular Poiseuille flow~\cite{alam} and how the 
confinement of the equilibrium hard-sphere fluid to restrictive one- 
and two-dimensional channels with soft interacting walls modifies its 
structure, dynamics, and entropy~\cite{mittal06,mittal07}. It has been 
shown recently~\cite{cesar1} that the origin of chaotic motion of two 
interacting particles in a one-dimensional box is due 
{\it double collisions} which occur {\it very close} to the hard walls. 
These double collisions occur when one particle is colliding with the 
1D wall and almost {\it simultaneously} collides with the other particle. 
As a consequence, the kind of motion 
generated close to the 1D walls is essential for the whole dynamics inside 
the billiard. Therefore we expect that the softness of the 1D walls will 
strongly affect the dynamics of the interacting particles.

In this contribution we generalize previous results~\cite{cesar1,cesar2} to 
the case of 1D soft walls. The equivalence between the two interacting 
particles in the 1D soft walls model with the motion of one particle inside 
a {\it soft} right triangular billiard is shown. In this right triangular
description the role of all important parameters from the problem becomes
clear. The interacting particles dynamics is studied by
varying the mass ratio $\gamma=m_{2}/m_{1}$ of the particles and the 
smoothness of the 1D wall potential. The reason to use the mass ratio as a 
dynamical parameter is related to the generation of new materials in the 
field of nanotechnology, where electrons may be confined inside a disk and 
can be affected by the surrounding material which composes the 
semiconductor~\cite{nano}. The composition of the surrounding material 
changes the effective mass between particles~\cite{poor,merkt}. We show here 
that a small ``degree of softness ($\sigma$)'' of the 1D walls, 
strongly decreases the mean values of the finite-time Lyapunov exponents (FTLEs).
The statistics of the distributions from the FTLEs has been studied in a number of
physical situations ranging from turbulent flows~\cite{wiggins} to Hamiltonian
dynamics (in many-particle system~\cite{berry}, and conservative
mappings~\cite{badii,lopes}). In this work we use the FTLEs distribution over
initial conditions, and the number of occurrences of the most probable FTLE,
which has been proposed~\cite{cesar1} as an efficient quantity to detect small
islands (dynamical traps) in phase space, to describe the qualitative and
quantitative appearance of regular and sticky motion as a function of $\sigma$. 

The paper is organized as follows. In Section~\ref{model} the model with soft
1D walls used in this contribution is presented. The description of this model in
the right triangular is given in Section \ref{triangular}. Section~\ref{lyapunov} 
shows a
systematic numerical study for the FTLEs, i.~e., the number of occurrences of 
the most probable FTLEs as a function of the mass ratios and of the
softness parameter. Poincaré Surfaces of Section (PSS) are used to show: 
a) The dynamics in the limit of hard 1D walls and b) the rise of sticky 
trajectories. We end with the conclusions in Section~\ref{conclusions}.

\section{The 1D soft walls model}
\label{model}

\begin{figure}[htb]
 \unitlength 1mm
 \begin{center}
 \includegraphics*[width=8cm,angle=0]{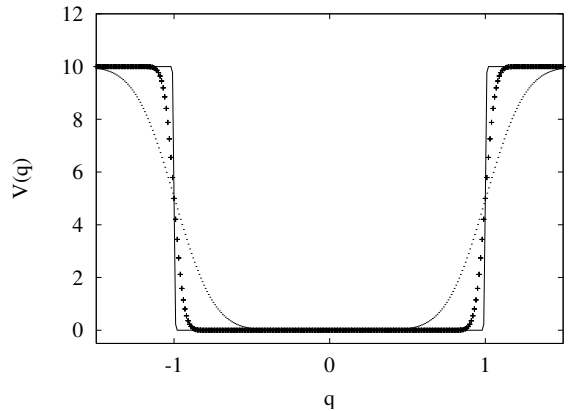}
 \end{center}
 \caption{Potential energy and corresponding forces for $F_0=10.0$ 
and $\sigma=5.0\times10^{-3}$ (filled line), $\sigma=5.0\times10^{-2}$ (cross 
points), $\sigma=9.0\times10^{-2}$ (dotted line).}
  \label{Vi}
  \end{figure}

In this section we introduce the model used for the investigation of 
two interacting particles inside the 1D billiard with soft walls. This model 
exhibit a continuous transition between soft and hard walls. The Hamiltonian 
considered is 

\begin{equation}
H=\sum_{i=1}^2\left[T_i+V_i(q_i)\right] + V_{{int}}=E,
\label{H}
\end{equation}
where $T_i=\frac{{{p}_{i}}^2}{2m_{i}}$ ($i=1,2$) is the kinetic energy of
particle $i$, $V_{{int}}=V_{0}/r$ is the Coulomb repulsion between
particles with $r=|q_1-q_2|$, and $V_i(q_i)$ is the potential energy from the 
1D soft walls, given by 

\begin{equation}
V_i(q_i)= \frac{F_0}{2}\left[erf \left( \frac{q_{i}-d_w}{\sigma \sqrt{2}} 
   \right)-erf \left( \frac{q_{i}+d_w}{\sigma \sqrt{2}} \right) \right] + F_0.
\label{Viqi}
\end{equation}
The first term on the right hand side of Eq.~(\ref{Viqi}) represents the soft 
1D wall located at $q=d_w$, while the second term represents the 1D soft wall 
located at $q=-d_w$. Here $\sigma$ is the parameter which quantifies the 
``softness'' of the walls and $F_0$ is the 1D walls intensity.
$F_0$ is added such that the total energy is positive. Figure \ref{Vi} 
shows the potential (\ref{Viqi}) for different values of the softness
parameter: $\sigma=5.0\times10^{-3}$ (filled line), $\sigma=5.0\times10^{-2}$ 
(cross points), $\sigma=9.0\times10^{-2}$ (dotted line). The 1D soft walls are
located at $q_w=\pm1$. As the softness 
parameter increases the walls become soft. An example of the 1D soft wall is 
shown by the dotted line in Fig.~\ref{Vi} for $\sigma=9.0\times10^{-2}$. As 
$\sigma$ approaches zero (see the filled line for $\sigma=5.0\times10^{-3}$) 
the walls looks very similar to the 1D hard wall. This limit will be called 
here as the 1D {\it quasi-hard} wall limit.

The corresponding left and right force of the walls on particle $i$ is
\begin{equation}
F_{i}=\frac{1}{\sqrt{2 \pi \sigma ^2}} e^{-\frac{(q_{i}+d_w)^2}{2 \sigma ^2}}
     -\frac{1}{\sqrt{2 \pi \sigma ^2}} e^{-\frac{(q_{i}-d_w)^2}{2 \sigma ^2}}.
\label{Fi}
\end{equation} 
The force of the walls has the Gaussian form, which in the limit
$\sigma\rightarrow 0$ approaches the $\delta$-function. In this limit the 
corresponding potentials $V_i(q_i)$ approach the hard walls.
Rescaling the time by $d\tau/dt=\sqrt{2E}$, the effective hamiltonian of the
1D soft walls model is

\begin{equation}
\tilde H=\frac{H}{2}=\sum_{i=1}^2\left[T_i+\tilde V_i(q_i)\right]+\tilde V_{int},
\label{HE}
\end{equation}
where  $V_{{int}}=\tilde V_{0}/r$ and $\tilde V_i(q_i)$ is given by
Eq.~(\ref{Viqi}) by using $\tilde F_0$ instead $F_0$. The scaled potential 
intensities are $\tilde V_0=\frac{V_0}{2E}$ and $\tilde F_0=\frac{F_0}{2E}$.
These scaled intensities show the role of the total energy on the dynamics.
Results of the present work, given for a combination of the scaled parameters, 
are valid for other energies keeping the scaled parameters constant.

The model with smooth walls presented here differs, to our knowledge, from 
all models used in the literature. In our case, we assume the 
$\delta$-function (in the limit $\sigma\rightarrow 0$) for the {\it force} 
instead for the potential, as usual. The reason to do this is simple: when 
we use the  $\delta$-function for the {\it potential}, then the corresponding 
force is not well defined. Our model makes it possible to study the classical 
dynamics continuously in the transition from smooth to hard walls. Beside that, 
hard walls discontinuities in numerical simulations may introduce errors at 
each wall collision.

\section{The 1D soft model and the soft triangular billiard}
\label{triangular}

It is well know \cite{casati99,mittag96,casati97} that the motion of three 
particles on a  frictionless ring with point-like interactions is equivalent 
to one particle moving freely inside the billiard, colliding elastically 
with the sides of the
triangle. Let us start with the Hamiltonian of the three particles on a ring, 
given by $H_B=\frac{p_1^2}{2m_1} + \frac{p_2^2}{2m_2}+\frac{p_3^2}{2m_3}$. 
We assume that  
the elastic collisions take place at $q_1=q_2$, $q_2=q_3$ and $q_3=q_1+1$. 
These collision points define the hard-wall sides of the triangle. 
Using the orthogonal transformation \cite{casati99}
($M=m_1+m_2+m_3$):

\begin{eqnarray}
q_1 &=& -\sqrt{\frac{m_3}{(m_1+m_2)M}}x - \frac{1}{m_1}
         \sqrt{\frac{m_1m_2}{(m_1+m_2)}}y+\frac{z}{\sqrt{M}},\cr
    & & \cr
q_2 &=& -\sqrt{\frac{m_3}{(m_1+m_2)M}}x +  
        \frac{1}{m_2}\sqrt{\frac{m_1 m_2}{(m_1+m_2)}}y+\frac{z}{\sqrt{M}},\cr
   & & \cr
q_3 &=&  \sqrt{\frac{(m_1+m_2)}{m_3M}}x + \frac{z}{\sqrt{M}},
\label{trans}
\end{eqnarray}
for the three particles on a ring, the resulting Hamiltonian is 
$H_B=\frac{1}{2} {\dot x}^2+ \frac{1}{2} {\dot y}^2+\frac{1}{2} {\dot z}^2$.
This is equivalent to the motion of one particle inside a triangular 
billiard with angles

\begin{equation}
\tan{\alpha}=\sqrt{\frac{m_2M}{m_1m_3}},\
\tan{\beta}=\sqrt{\frac{m_1M}{m_2m_3}}, \
\tan{\eta}=\sqrt{\frac{m_3M}{m_1m_2}}.
\nonumber
\label{angles}
\end{equation}
The point-like collision between particles $1$ and $2$ defines one 
side of the triangle at $q_1-q_2=0$, and the collision of 
these particles with particle $3$ defines the other two sides 
of the same triangle. For $m_3\rightarrow\infty$ ($\eta=\pi/2$) 
we get the right triangular billiard which corresponds to the 
motion of two particles $m_1$ and $m_2$ moving inside the 1D
box with hard walls.  In this case the interaction 
between particles $1$ and $2$ is the point-like collision and the 
fixed particle $3$ plays the role of the 1D hard-wall, kept fixed 
at $q_3$. These collisions with the 1D fixed hard-wall can be 
represented by delta functions. However, the corresponding equations 
of motion are not well defined. Therefore, to describe analytically 
such problems we include in $H_B$ the soft interactions between particles
which, in a given limit, are expected to describe the point-like 
collisions and the 1D box hard-walls. 

Firstly we assume that the interaction between particles $1$ and $2$ is
the Coulomb repulsion  $V_{12}=V_{int}=V_0/|q_1-q_2|$. In order to give 
an idea of the transition to hard-walls in the triangle, we use the Yukawa 
potential $V_{12}=V_0e^{(-|q_1-q_2|/\nu)}/|q_1-q_2|$. The interaction between 
particles $1$ and $2$ depends only on the relative positions of both 
particles and $\nu>0$ is a parameter which allows to change the interaction 
range from the smooth potential. For $\nu\rightarrow0$ the limit of short 
interactions (point-like collision) can be approached. Using the orthogonal 
transformation (\ref{trans}) the above interaction is written as 
$V_{12}(|y|)=\sqrt{\mu_{12}}V_0e^{(-\frac{|y|}{\nu\sqrt{\mu_{12}}})}/|y|)$, 
where $\mu_{12}=m_1m_2/(m_1+m_2)$ is the reduced mass between particles $1$ 
and $2$. Using $V_{12}(|y|)$, the one side of the triangle located at $y=0$ 
is now soft, as will be shown numerically later.

Secondly, we assume that particles $1$ and $2$ interact smoothly 
with particle $3$. In order to study the 1D box case, the location of 
particle $3$ is kept fixed at $q_3=\pm d_w$ and $m_3\rightarrow\infty$. 
In this way particle $3$ will play the role of the 1D {\it soft}-walls. 
Here we consider the soft interactions $V_{i3}(x,y)=V_i(q_i)$ given
by Eq.~(\ref{Viqi}). Due to the soft interaction with particle $3$, particles 
$1$ and $2$ can, in the ring description, interact with particle $3$ on both 
sides, left ($q_3=-d_w$) and right ($q_3=+d_w$). In the limit 
$m_3\rightarrow\infty$ the orthogonal transformation (\ref{trans}) is then 
reduced to the ($x,y$) plane

\begin{eqnarray}
q_1 &=& -\frac{\sqrt{\mu_{12}}}{m_1}\left(y + \frac{m_1}{\sqrt{m_1m_2}}x\right),\cr
    & & \label{tranS}\\
q_2 &=& \frac{\sqrt{\mu_{12}}}{m_2}\left(y - \frac{m_2}{\sqrt{m_1m_2}}x\right).\cr
\nonumber
\end{eqnarray}
Rescaling the time by $d\tau/dt=\sqrt{2E}$ and using
the reduced orthogonal transformation (\ref{tranS}) we obtain, 
after straightforward calculation, the final scaled Hamiltonian 
$\tilde H_B= H_B/(2E)=1/2$ in the right triangular description 
[using ($\tilde x, \tilde y$) $\rightarrow$ ($x,y$)]

\begin{equation}
\tilde H_B=\frac{{\dot x}^2}{2}+ \frac{{\dot y}^2}{2}+\tilde V_{12}(|y|)+
\sum_{i=1}^2\tilde V_{i3}(x,y),
\label{HF}
\end{equation}
where 

\begin{eqnarray}
\tilde V_{12}(|y|)& = &  
  \tilde 
 V_0\frac{e^{\left(-\frac{|y|}{\sqrt{\mu_{12}}\nu} \right)}}{\left|y\right|},\cr
        & & \cr
\tilde V_{i3}(x,y)& = &\tilde F_0+\cr
& & \left.\right.\cr
 \frac{\tilde F_0}{2}& &\hspace{-0.7cm}\left\{erf 
    \left[ \frac{\sqrt{2\mu_{12}}}
   {2m_i\sigma}\left(y+\frac{(-1)^{i+1}m_i}{\sqrt{m_1m_2}}x\right)-
    \frac{\sqrt{2}d_w}{2\sigma} \right]-\right.\cr
& & \left.\right.\cr 
& & \left.\right.\cr 
& & \hspace{-0.8cm}\left.
erf \left[\frac{\sqrt{2\mu_{12}}}{2m_i\sigma}
\left(y+\frac{(-1)^{i+1}m_i}{\sqrt{m_1m_2}}x\right)+\frac{\sqrt{2}d_w}{2\sigma} \right]  
\right\}.
\nonumber
\label{HscaledNOS}
\end{eqnarray}
The total energy of the scaled problem is $\tilde H_B=1/2$
and the effect of the real energy can be seen in the scaled 
potential intensities
$\tilde V_0=\frac{V_0\sqrt{\mu_{12}}}{2E}$ and
$\tilde F_0=\frac{F_0}{2E}$.
In the above Hamiltonian we observe the role of the crucial quantities 
of the problem, i.~e.~, masses $m_1,m_2$, the softness ($\nu$) of the 
interaction between particles $1$ and $2$, the softness
($\sigma$) of the 1D {\it soft}-walls, and the size $d_w$ of the 1D soft 
walls billiard.
 \begin{figure}[htb]
 \unitlength 1mm
 \begin{center}
 \includegraphics*[width=7.2cm,angle=0]{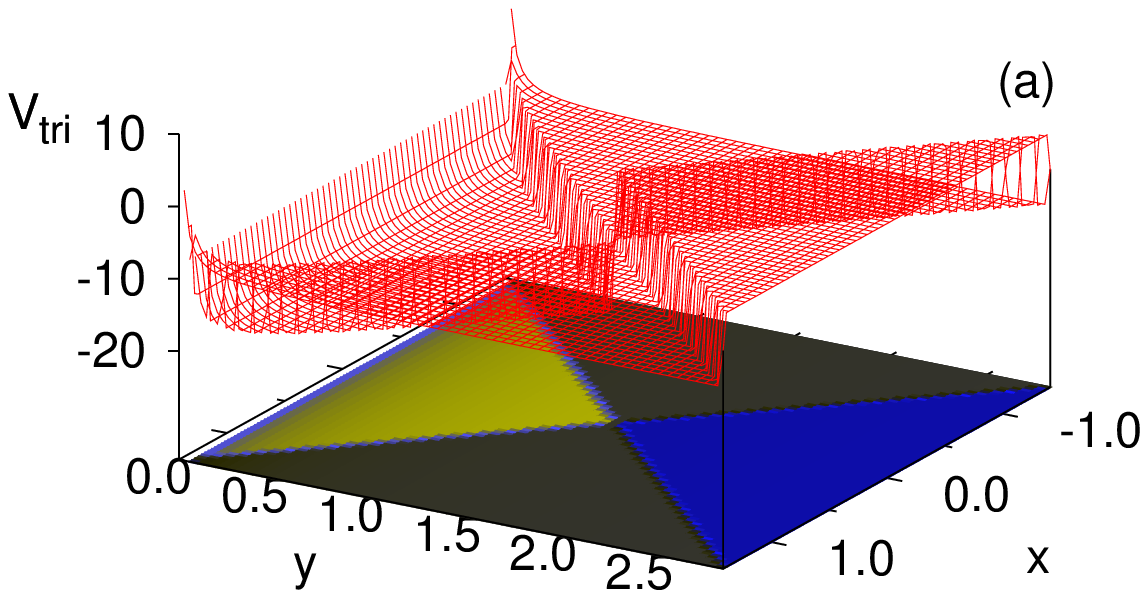}
 \includegraphics*[width=7.2cm,angle=0]{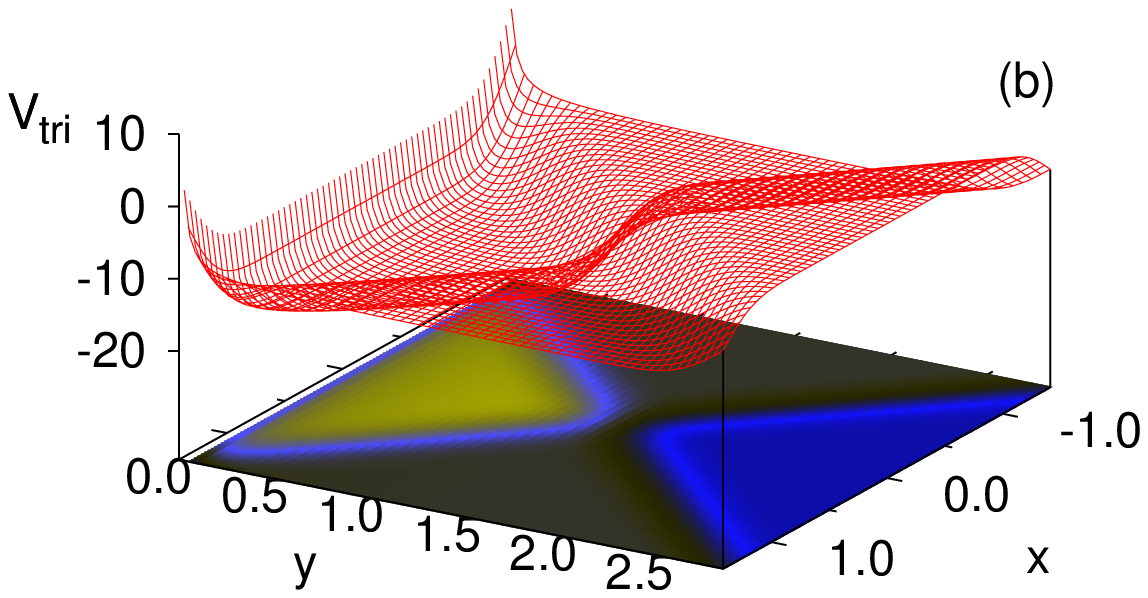}
 \includegraphics*[width=7.2cm,angle=0]{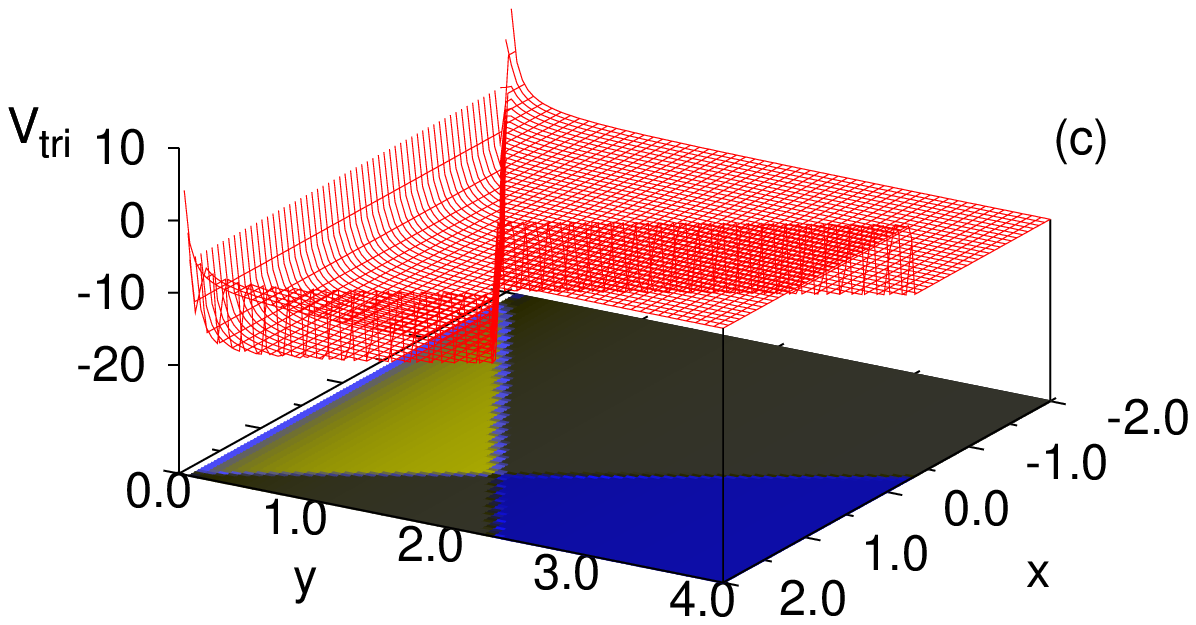}
 \includegraphics*[width=7.2cm,angle=0]{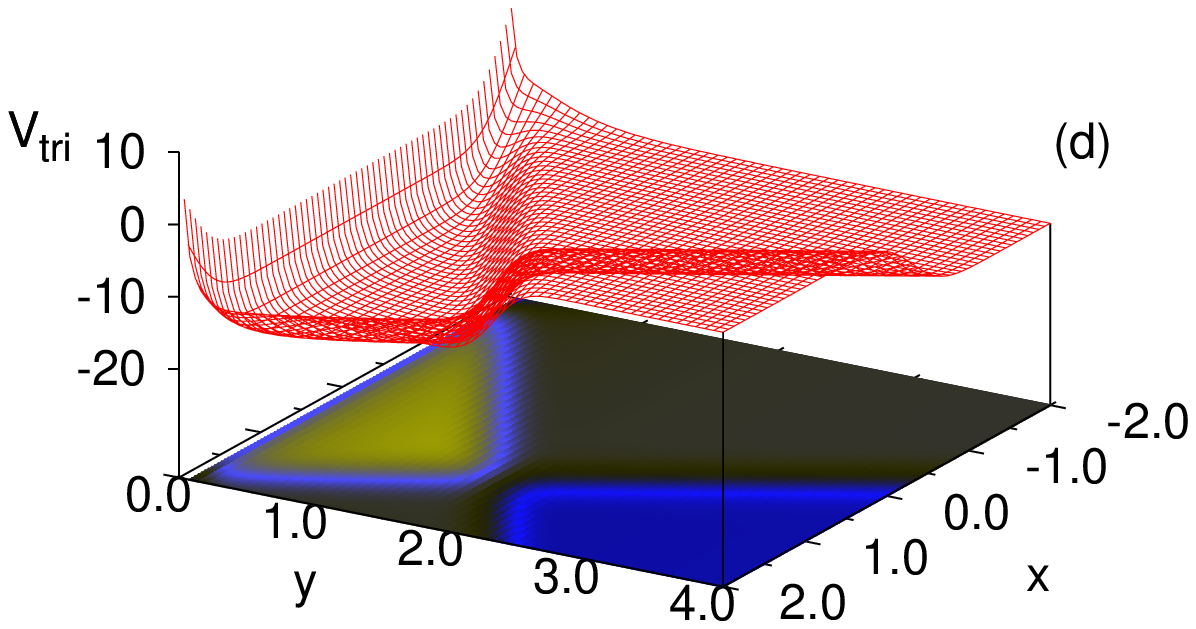}
 \end{center}
 \caption{(Color online) Scaled triangular potential energy ($V_{\mbox{tri}}$)
  from the Hamiltonian (\ref{HF}) with   $\tilde F_0=10.0$, $\tilde V_0=0.5$ 
  and $\sigma=5.0\times10^{-3}$ in
     (a) and (c), and $\sigma=2.0\times10^{-1}$ in (b) and (d). For 
     (a) and (b) we have $m_1=m_2=1$, and for (c) and (d) we have
     $m_1=3$ and $m_2=1$.}
  \label{Vi3d}
  \end{figure}

Figure \ref{Vi3d} shows the potential energy from (\ref{HF}) for
two values of the smoothness $\sigma$, and for two values of the 
masses $m_1$ and $m_2$. We assume that $\tilde V_{12}=\tilde V_0/|y|$.
First observation is that in the ($x,y$) plane the potential energy has 
a right triangle form, as expected. The left side of the triangle ($y=0$) 
has a smooth form due to $V_{12}(|y|)$, which is the long Coulomb repulsion
between particles $1$ and $2$. The other two sides of the triangle (with 
internal angle $\eta=\pi/2$) can be soft or not, depending on $\sigma$. 
These two sides are equivalent to the 1D soft walls. For 
$\sigma=5\times10^{-3}$ and $m_1=m_2=1$ we nicely see in Fig.~\ref{Vi3d}(a)
that these two sides approach to hard-walls, while for 
$\sigma=0.2$ these two sides are very soft, as can be seen in 
Fig.~\ref{Vi3d}(b). Figures \ref{Vi3d}(c)-(d) show the potential energy 
for the right triangle when the masses are changed ($m_1=1$ and $m_2=3$).  
Fig.~\ref{Vi3d}(c) for $\sigma=5\times10^{-3}$ and 
Fig.~\ref{Vi3d}(d) for $\sigma=0.2$. Above results show us that the
motion of two particle inside the 1D soft walls is equivalent to the 
motion of one particle inside the right triangle with soft walls.

Some limiting situations can promptly be seen from Hamiltonian 
(\ref{HF}): (a) When $F_0=0$ (no 1D walls) the two sides with internal
angle $\eta=\pi/2$ of the triangle disappear and just the soft wall 
$\tilde V_{12}(|y|)$ remains. The Hamiltonian is separable and integrable 
since the $x$ dependence in the potential energy disappears. This 
means that for two interacting particle (through relative coordinates)
which are not bounded inside walls, the problem is regular, as expected; 
(b) The kind of coupling between coordinates $x$ and $y$ depends on the 
form of the interaction used for $\tilde V_{i3}(x,y)$. Therefore, these two 
sides
(or the 1D soft-walls) determine if the dynamics inside the right triangle 
(or inside the 1D box) is chaotic/regular; (c) When $m_2\rightarrow\infty$ 
(or $m_1\rightarrow\infty$) the $x$ dependence in the argument of the error 
function (in $\tilde V_{i3}(x,y)$) can be neglected when compared to the $y$ 
dependence, and the Hamiltonian (\ref{HF}) is separable again. This is also
an expected result, since when particle $2$ (or $1$) is too heavy,
the motion should be regular again; (d) increasing the billiard size
$d_w$, the 1D soft walls get apart. The same effect is obtained by 
decreasing the softness $\sigma$; (d) the last limit which we would like to 
mention, and the most interesting one, is the hard-wall billiard case,
where $\sigma\rightarrow0$ and $\nu\rightarrow0$ (and 
$\tilde V_0\rightarrow\infty$). In this limit the whole 
dynamics can be explained in terms of rational/irrational values of 
$\alpha/\pi$ \cite{casati,casati97}. Using the Yukawa interaction in the 
1D hard walls billiard, we approached this limit numerically ($\nu=0.1$) 
in a previous work 
\cite{cesar1}. In \cite{cesar1} it was also shown analytically the 
influence of the interaction between particles $1$ and $2$ to generate 
positive Lyapunov exponents. The softness of this interaction is 
essential to generate positive Lyapunov exponents, and it is possible 
to see that for the point-like collisions the probability to 
obtain positive Lyapunov exponents goes to zero. To study this limit 
analytically, the interaction potentials between all particles must 
be chosen appropriately, and it is the subject of a future work.
In this context it would be interesting to use the methodologies 
developed in \cite{ana,turaev98,turaev99,turaev03} to study the
limit $\sigma\rightarrow0$.

We finish this section by mentioning that the Hamiltonian (\ref{HF})
corresponds to the motion of one particle inside the right triangle
suffering soft collisions at the walls. Although this is not, strictly 
speaking, a billiard motion as in the original sense (free particle inside 
a table), we will refer to it as the particle inside a soft right triangular 
billiard, due to the analogy  shown in this section.

\section{Results}
\label{lyapunov}

We investigate the non-linear behavior inside the 1D soft billiard 
by determining the distribution $P(\Lambda_{t},\gamma)$ of the finite-time 
largest Lyapunov exponents $\Lambda_{t}$ as a function of the mass ratio
between particles $\gamma=m_2/m_1$. This investigation was done previously for 
the 1D hard-wall case \cite{cesar1,cesar2} and the whole dynamics depends 
strongly on $\gamma$. The FTLE is obtained by integrating two closed 
trajectories, 
computing the local LE after a time $\tau=0.1$ and making an average over all
the local LEs. The time $\tau$ is chosen for the better convergence of the LE.
We used the fourth-order Runge Kutta method with variable steps. The energy is 
conserved in all simulations by around $10^{-6}$. For quasi-integrable system, 
the presence of broken KAM curves inside  the chaotic sea in phase space, leads 
to trapping and `sticky' trajectories~\cite{zas}, affecting the convergence in 
the determination of the FTLEs,  which depend now on the initial conditions. 
On the other hand, it implies that the distribution $P(\Lambda_{t},\gamma)$,
calculated over many initial conditions, contains information about the amount 
of regular motion (and trapped trajectories) in phase space
\cite{politi,grebogi,titov,cesar1,falcioni,steve}.
For ergodic system and for infinite times, the Lyapunov exponents do 
not depend on initial conditions~\cite{oseledec}.  

\subsection{The 1D quasi-hard limit ($\sigma=5\times 10^{-3}$)}
\label{quasi}

\begin{figure}[htb]
 \unitlength 1mm
 \begin{center}
 \includegraphics*[width=9cm,angle=0]{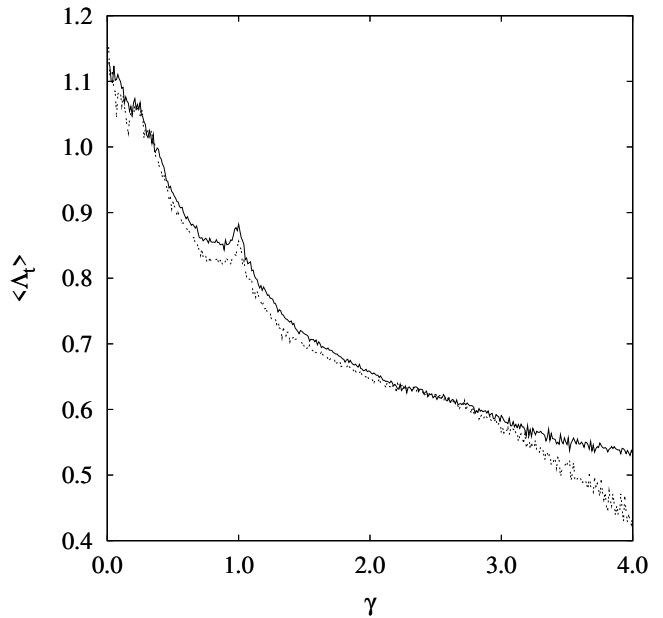}
 \end{center}
 \caption{Mean values of the finite-time largest Lyapunov exponent
 calculated over $200$ trajectories up to time $t = 10^{4}$ and at 
 scaled energy $\tilde E=0.5$, $ \tilde V_{0}=1.0$, $\tilde F_0=10.0$, 
  for $\sigma=5\times 10^{-3}$ (dashed line). This is the 1D quasi-hard wall 
  limit.  For comparison, the full line shows results obtained for the 1D hard
  walls~\cite{cesar2}. For each trajectory the largest FTLE is evaluated over
  $10^5$ samples.}
  \label{mean}
  \end{figure}

Figure \ref{mean} shows (dashed line) the mean FTLE
$\langle\Lambda_{t}\rangle$ as a function of the mass ratio $\gamma$ for
$\sigma =5\times 10^{-3}$. This is the case of very small values of $\sigma$ 
(1D quasi-hard wall limit) and the 1D soft walls look very similar to the 
1D hard-wall potential (see Fig.~\ref{Vi}). 
The mean FTLE decreases from roughly $\sim 1.15$ 
to $0.54$ in the whole mass ratio interval. This means that the dynamics is 
getting more and more regular, as expected, since  $\gamma\rightarrow\infty$ 
constitutes an integrable limit with the heavy particle at rest. However, 
at $\gamma\sim1.0$ the mean FTLE increases, showing that the symmetry for equal 
masses increases the amount of irregular motion.  Figure \ref{mean} also shows 
(full line) results for the case of hard 1D walls, published previously 
\cite{cesar2}. In that case, and therefore in the case observed here, the peak at
$\gamma=1.0$ is the consequence of the resonance observed in the limit of
hard-point collisions between particles.
The qualitative behavior of the FTLEs could be explained in~\cite{cesar2} 
with the help of some special periodic orbits from the Gauss map. For the 
purpose of this work we can observe that both curves in Fig.~\ref{mean} 
are in good agreement. This confirms that the model of 1D soft walls presented 
here reproduces correctly, in the limit $\sigma\rightarrow0$, the 1D hard wall 
case. However, small differences can be observed in Fig.~\ref{mean} for 
$\gamma\lesssim0.4, \gamma\sim0.96$ and $\gamma \gtrsim 3.0$. These 
differences  will be explained later, where we also show some Poincar\'e 
Surfaces of Section~(PSS) and discuss more details of the whole dynamics 
for different values of the mass ratio.

\begin{figure}[htb]
 \unitlength 1mm
 \begin{center}
 \includegraphics*[width=8.5cm,angle=0]{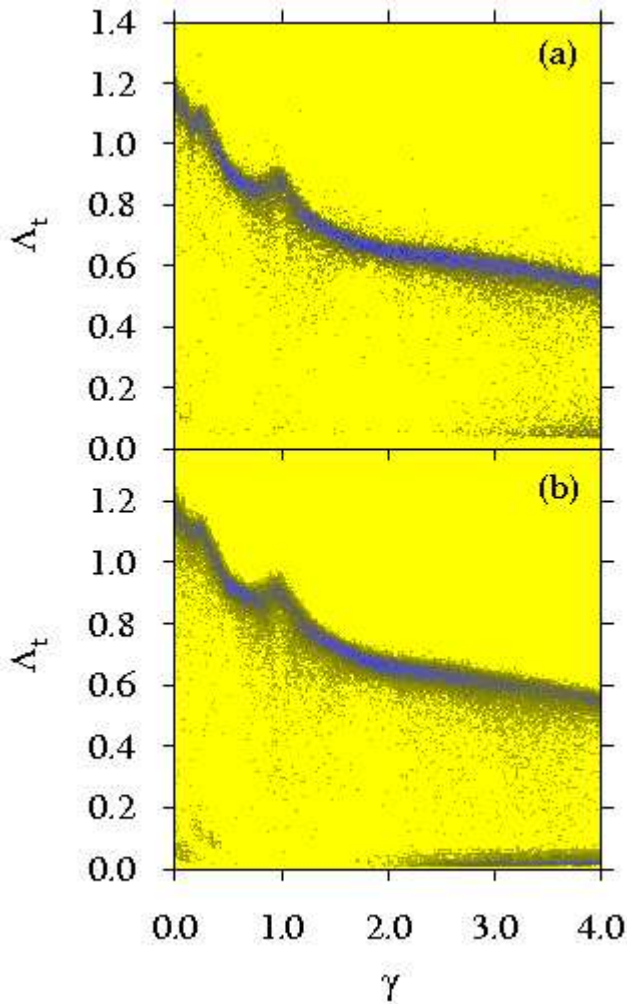}
 \end{center}
 \caption{(Color online)Finite-time distribution of the largest Lyapunov 
 exponent $P(\Lambda_{t},\gamma)$ calculated over $200$ trajectories up to 
 time $t = 10^4$, for (a) 1D quasi-hard wall limit $\sigma= 5\times 10^{-3}$, 
 and (b) 1D hard wall limit from~\cite{cesar2}. With increasing 
$P(\Lambda_{t},\gamma)$ the color changes from light to dark (white over
   yellow and blue to black).} 
  \label{dist}
  \end{figure}

To analyze deeper the effect of the smooth potential on the particles
dynamics, Fig.~\ref{dist} shows the finite-time distribution of the largest 
Lyapunov exponent, $P(\Lambda_{t},\gamma)$, for the (a) 1D quasi-hard wall limit 
$\sigma=5\times 10^{-3}$ and (b) the 1D hard wall case from~\cite{cesar2}. The
gray points below the main curve are related to chaotic  trajectories which 
were trapped for a while close to regular islands. Since both figures are
quite similar, the 1D quasi-hard limit also represents adequately the 1D hard 
wall case concerning trapped trajectories.
An interesting feature in  Fig.~\ref{dist} is the change of the width
of $P(\Lambda_{t},\gamma)$ around the number of occurrences of the most 
probable $\Lambda^{p}_{t}$ defined through~\cite{cesar1}

\begin{equation}
    \label{probable}
   \left. \frac{\partial
P(\Lambda_{t},\gamma)}{\partial
\Lambda_{t}}\right|_{\Lambda_{t}=\Lambda_{t}^{p}}\,=0.
\end{equation}
For mass ratios close to $\gamma \sim 1.0$, for example, many initial 
condition lead to different values of $\Lambda_{t}$.  In this region,
$\Lambda^{p}_{t}$ has a minimum as a function of $\gamma$, 
which is a clear demonstration of the presence of `sticky' trajectories.

 \begin{figure}[htb]
 \unitlength 1mm
 \begin{center}
 \includegraphics*[width=8cm,angle=0]{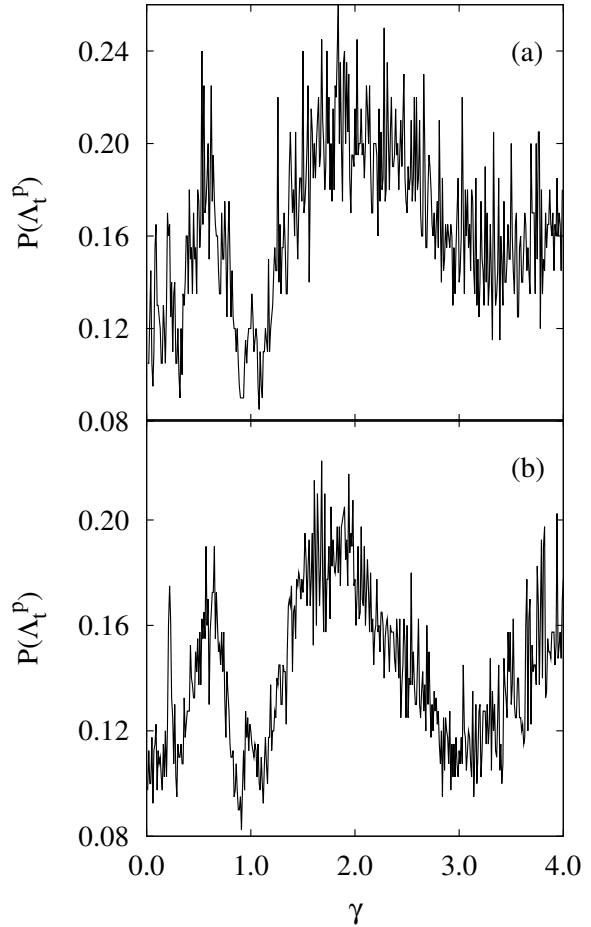}
 \end{center}
\caption{Normalized values of the most probable Lyapunov exponent 
  $\Lambda_{t}^{p}$ for (a) 1D quasi-hard case $\sigma= 5\times 10^{-3}$ 
 and (b) 1D hard wall case from~\cite{cesar2}.}
  \label{max}
  \end{figure}
To determine the amount of `sticky' and regular trajectories in phase space 
we follow  $\Lambda^{p}_{t}$ as a function of the mass ratio $\gamma$. In fact,
we follow $P(\Lambda^{p}_{t})$, which is the normalized number of occurrences 
of the most probable LE, or the probability to obtain $\Lambda^{p}_{t}$. This 
is shown in Fig.~\ref{max}. When  $P(\Lambda^{p}_{t})$ is large, a large 
fraction of initial conditions lead to the same $\Lambda_{t}$ and trapped 
trajectories are rare. For example, the maximum of  $P(\Lambda^{p}_{t})$
in Fig.~\ref{max} (a) [and (b)] close to $\gamma \sim 1.8$, is the region 
in Fig.~\ref{dist}(a) where gray points below the main curve are rare. On the
other hand, close to $\gamma\sim0.96$ we have a minimum in  Fig.~\ref{max} (a) 
[and (b)], which is the consequence of the large dispersion around
$\gamma\sim0.96$ in Fig.~\ref{dist}(a). Again the quasi-hard limit 
[Fig.~\ref{max} (a)] and hard wall [Fig.~\ref{max} (b)] 
agree very well. The fast variation of  $P(\Lambda^{p}_{t})$ is due to 
statistical fluctuations in its determination over initial conditions.  

\begin{figure}[htb]
\begin{center}
 \includegraphics*[width=4.3cm,angle=0]{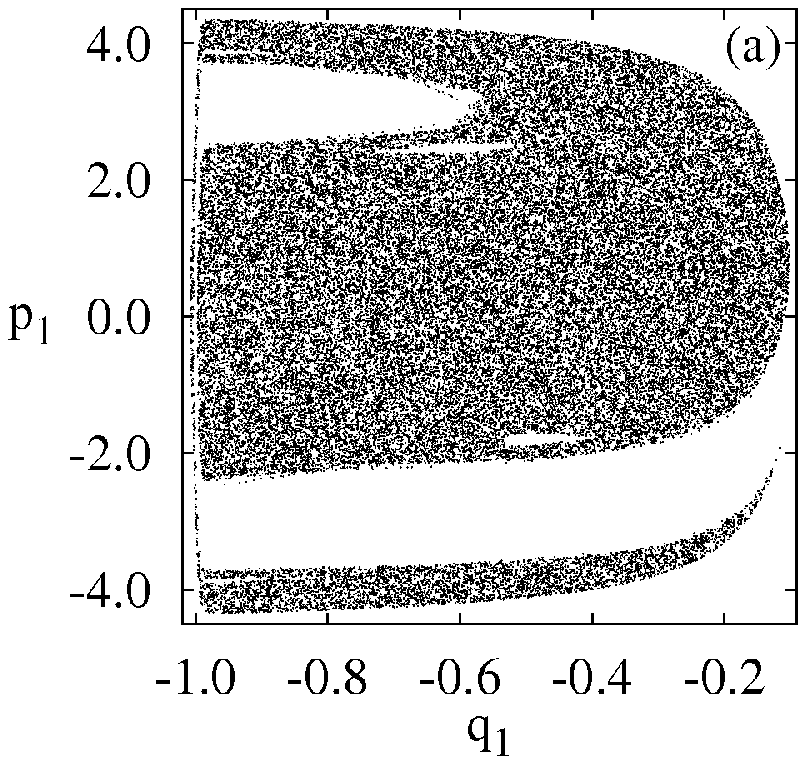}
 \includegraphics*[width=4.2cm,angle=0]{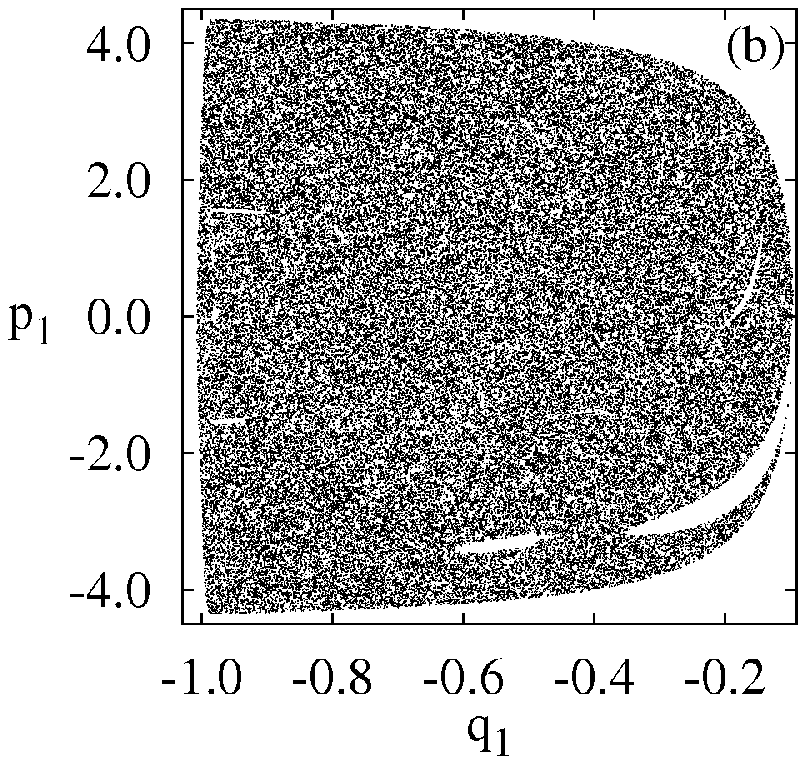}
 \includegraphics*[width=4.2cm,angle=0]{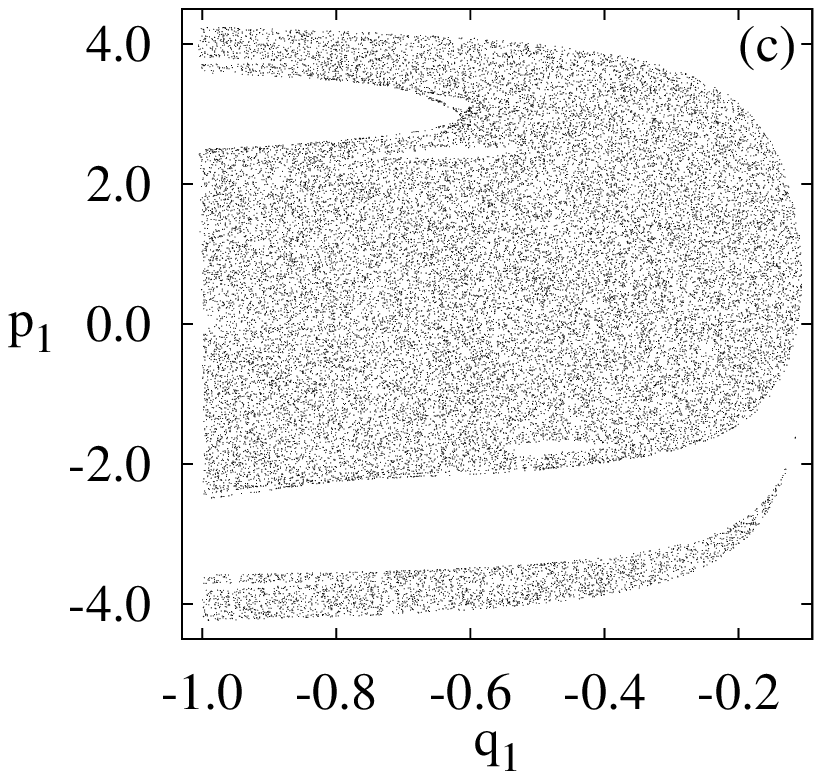}
 \includegraphics*[width=4.2cm,angle=0]{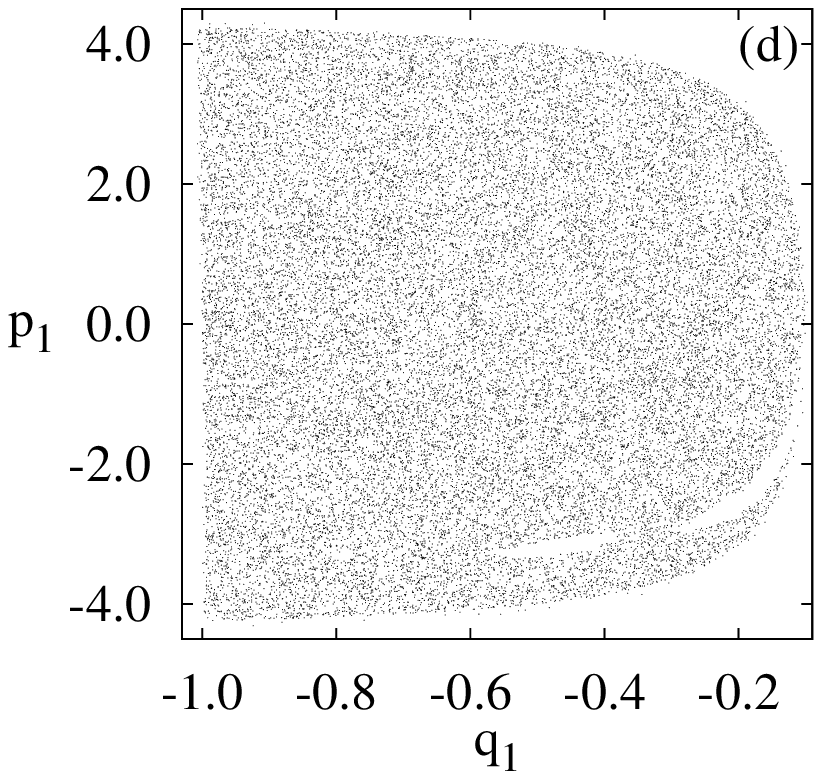} 
 \leavevmode
\end{center}
\caption{Poincar\'e surfaces of section for particle $1$ and (a)-(b) 1D quasi-hard
  limit and (c)-(d) 1D hard wall case. Figures (a) and (c) for $\gamma=1.0$, and
  (b) and (d) for $\gamma=1.8$.}
  \label{PSS1.0-1.8}
  \end{figure}

Figure \ref{PSS1.0-1.8} shows the PSS for $\gamma=1.0,1.8$ and compares
the 1D quasi-hard limit [(a)-(b)] with the 1D hard wall case [(c)-(d)]. 
Both cases are alike. The PSS is constructed in the following way: each 
time particle $2$ is located at the origin, and $p_2>0.0$, then the point
($q_1,p_1$) is recorded. 
Since this work is focused on the dynamics inside the realistic
1D box problem, we have chosen to present results of the PSS in the 
1D box coordinates, instead in the right triangular coordinates ($x,y$).
In fact, we have checked some PSS in the ($x,y$) coordinates and no
additional relevant informations for the present work were obtained.
It can be observed in Fig.~\ref{PSS1.0-1.8} that for $\gamma=1.8$ the PSS 
is almost filled with
the chaotic trajectory, while for $\gamma=1.0$ some regular islands
and forbidden trajectories appear which induce the trapping trajectories
(These trapped trajectories can not be seen in the resolution used 
in Fig.~\ref{PSS1.0-1.8}). Therefore, in addition to the mean FTLE and
the number of occurrences of the FTLE, also in the PSSs the quasi-hard 
limit and the hard 1D wall case are almost identical, showing that our 
soft model correctly describes the hard 1D wall limit 
$\sigma\rightarrow0$. This will change drastically at next, when the 
softness parameter increases.

\subsection{The 1D soft case $\sigma=5\times 10^{-2}$}

\begin{figure}[htb]
 \unitlength 1mm
 \begin{center}
 \includegraphics*[width=8cm,angle=0]{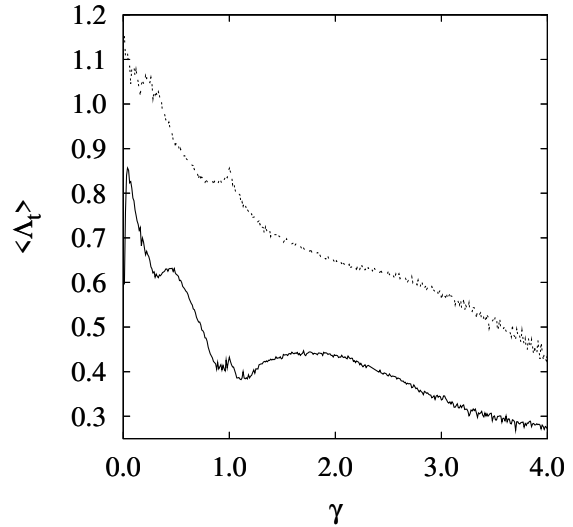}
 \end{center}
 \caption{Mean values of the finite-time largest Lyapunov exponent
 calculated over $400$ trajectories up to time $t = 10^{4}$ and at 
 scaled energy $\tilde E=0.5$, $ \tilde V_{0}=1.0$, $\tilde F_0=10$  for 
 the 1D soft wall with 
 $\sigma=5\times 10^{-2}$ (full line). For comparison, the full line shows 
 results obtained for the quase-hard 1D wall from Fig.~\ref{mean}.
 For each trajectory the largest FTLE is evaluated over $10^5$ samples.}
  \label{mean2}
  \end{figure}

Finally we discuss the case of 1D soft walls. The form of the 1D wall
potential is shown in Fig.~\ref{Vi} (see cross points for 
$\sigma=5\times 10^{-2}$). Although the value of $\sigma$ is small, the 
effect on the particles dynamics is astonishing. Figure \ref{mean2} 
shows the mean values of the finite-time largest Lyapunov exponents 
for $\sigma=5\times 10^{-2}$ (full line) compared with the quasi-hard 
1D wall limit (dashed line) from Fig.~\ref{mean}. The effect of the small 
soft potential is remarkable. The mean FTLEs decrease around $50\%$ and the 
peak observed close to $\gamma=1.0$ (full line) is now a minimum. A 
new minimum also appears close to $\gamma\sim 0.40$.

\begin{figure}[htb]
 \unitlength 1mm
 \begin{center}
 \includegraphics*[width=8cm,angle=0]{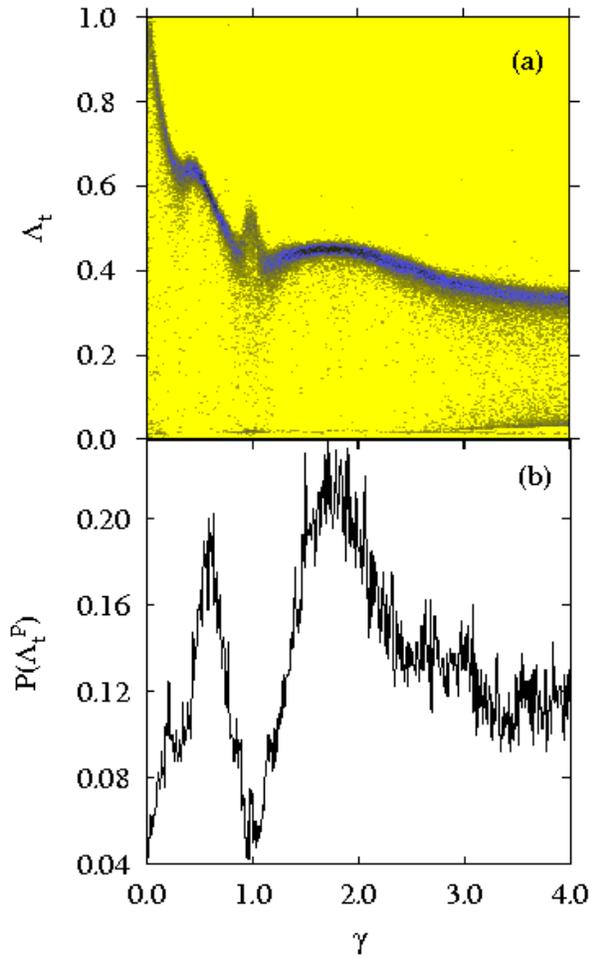}
\hspace*{-0.6cm} 
 \end{center}
 \caption{(Color online) (a) Finite-time distribution of the largest 
 Lyapunov exponent
 $P(\Lambda_{t},\gamma)$ calculated over $400$ trajectories up to time
 $t = 10^4$,  $\sigma= 5\times 10^{-2}$.  With increasing
 $P(\Lambda_{t},\gamma)$ the color changes from light to dark (white
 over yellow and blue to black); (b) Corresponding normalized number of
 occurrences of the most probable Lyapunov exponent $\Lambda_{t}^{p}$.}
  \label{dist2}
  \end{figure}

To understand better what happens, Fig.~\ref{dist2}(a) shows the finite-time 
distribution of the largest Lyapunov exponent for this case. Clearly we see 
that, although the peak close to $\gamma=1.0$ still exist, the dispersion 
around the most probable FTLE is very large. This large dispersion affects 
the mean Lyapunov exponents from Fig.~\ref{mean} and a minimum occurs at 
$\gamma=1.0$. This strong dispersion around $\gamma=1.0$ is also confirmed
by the accentuated minimum of the number of occurrences of the most probable 
FTLE plotted in Fig.~\ref{dist2}(b).
A similar effect, but with smaller intensity occurs for $\gamma\sim0.4$. 
The mean FTLEs increase as $\gamma$ decreases, the amount of gray points 
below the main curve increases and the number of occurrences of the most 
probable FTLE also decreases. 

Many other values of the mass ratio $\gamma$ could be discussed to 
analyze the appearance of island in phase space. Instead of doing
so we would like to show some PSS in order to discuss the dynamics
of the interacting particles in the 1D soft walls. We start showing the
case of $\gamma=1.0$, for which many trapped trajectories and islands 
are expected.
 \begin{figure*}[htb]
 \begin{center}
 \includegraphics*[width=4cm,angle=0]{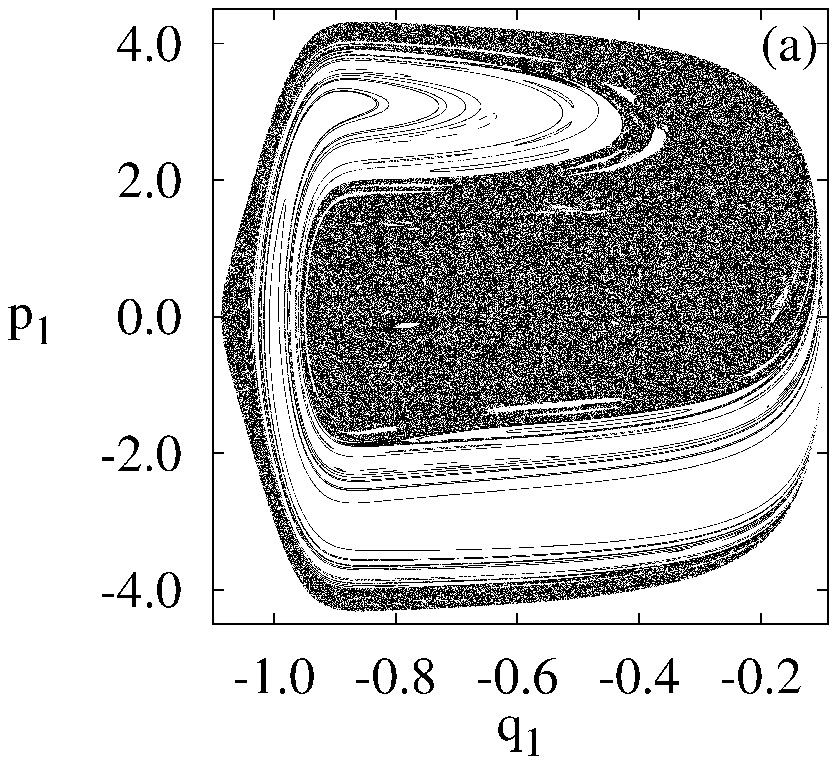}
 \includegraphics*[width=4cm,angle=0]{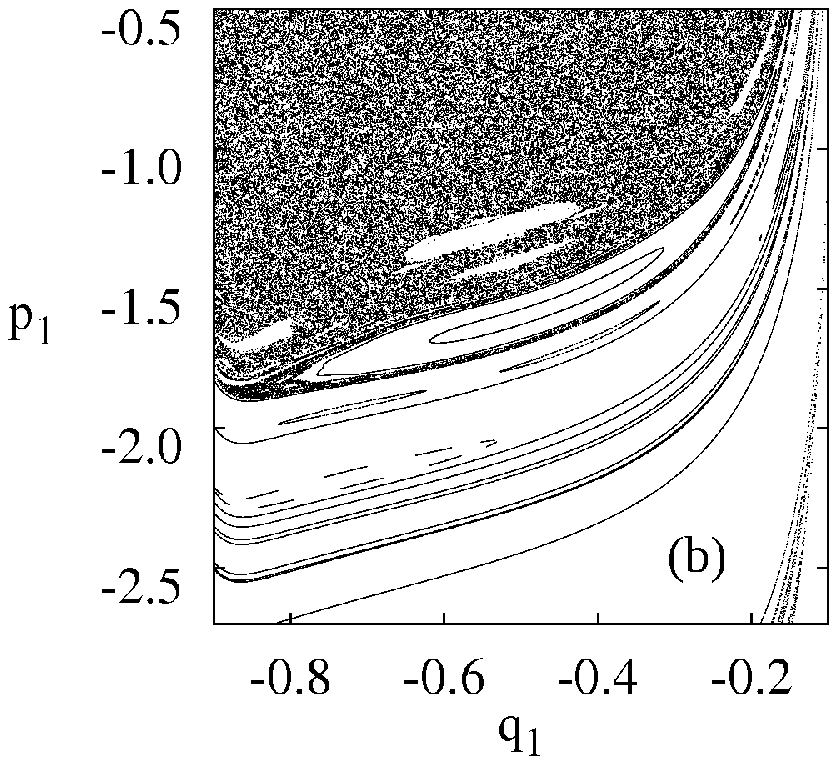}
 \includegraphics*[width=4cm,angle=0]{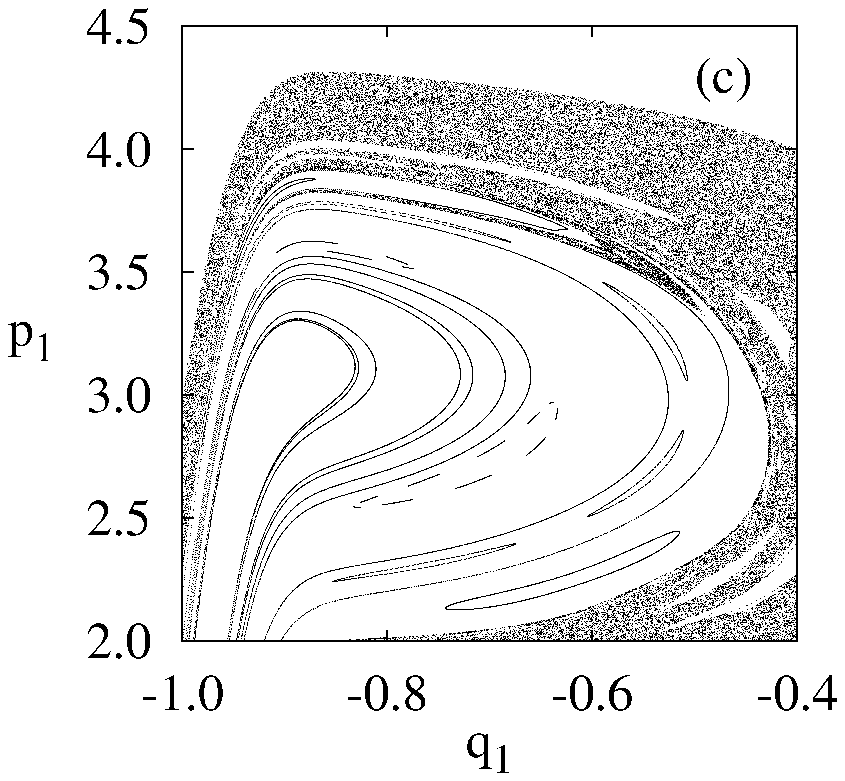}
 \includegraphics*[width=4cm,angle=0]{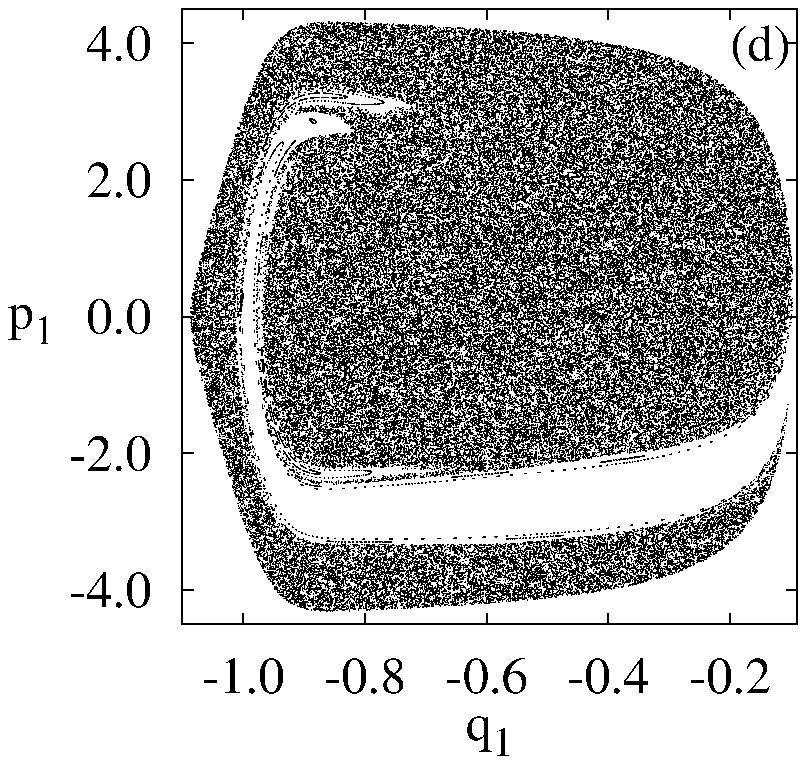} 
 \end{center}
\caption{Poincar\'e surfaces of section for $\sigma=5\times 10^{-2}$ and for 
  (a) $\gamma=1.0$, and their magnification (b) and (c). Figure (d) is for 
  $\gamma=0.8$.}
  \label{PSS1.0}
  \end{figure*}
Figure \ref{PSS1.0}(a)-(c) shows the corresponding PSS. Since for the 
PSS we used the condition $p_2>0.0$, particle $2$ is always moving to 
the right at $q_2=0$. Because the interaction between particles is 
repulsive, we expect that particle $1$ crosses the PSS more often for 
$p_1<0.0$ than for $p_1>0.0$. This physical effect is the origin of the 
asymmetry in Fig.~\ref{PSS1.0}(a) observed in $p_1$. Additionally, due
to $p_2>0.0$, trajectories on the 
PSS with negative momentum ($p_1<0.0$) are less affected by particle $2$ 
than trajectories with $p_1>0.0$. Since the chaotic motion results from 
the simultaneous effect of the 1D walls and the mutual interactions, we 
conclude that trajectories with $p_1<0.0$ should be more regular. 
Therefore we expect more island for $p_1<0.0$ than for  $p_1>0.0$
[See Fig.~\ref{PSS1.0}(a)].
Observe that for  $p_1<0.0$ the islands extend themselves from 
$q_1\sim -1.0$ until $q_1\sim -0.1$, which is very close to particle $2$, 
located at $q_2=0$. For $p_1>0$ the interval of islands goes from 
$q_1\sim -1.0$ until $q_1\sim -0.4$, very far from particle $2$.
Close to the left 1D wall ($q_1\sim -1.0$), the momentum of particle 
$1$ remains in the interval $-1.0\lesssim p_1\lesssim 1.0$, while close 
to particle $2$ ($q_1\sim -0.1$) we have chaotic trajectories with 
essentially only positive momenta ($0.0\lesssim p_1\lesssim 1.0$) due the 
repulsion of particle $2$. For $-1.0\lesssim p_1\lesssim 0.0$ we have
essentially regular trajectories since both particles are moving apart.
The main island in Fig.~\ref{PSS1.0}(a) is related to the following 
{\it regular trajectory}: particle $1$ is moving to the left with a small 
momentum which does not allow it to ``penetrate'' very much the 1D soft wall. 
It ``collides'' with the 1D soft wall and simultaneously is affected by the 
interaction force with magnitude comparable to the magnitude of the 1D wall 
force. This is called here as the {\it low-energy double collision}, 
which in this example occurs between particle $1$ (low-momentum), 
particle $2$ and the 1D soft wall. After colliding with the left 1D wall, 
particle $1$ moves to the right with a small momentum and can not 
approach particle $2$ very much due to the long Coulomb repulsion. 
Subsequently it changes its direction and goes back to the left 1D wall.

Another interesting property occurs very close to the 
``top of the 1D soft wall potential''. We call the readers to attention 
that ``top of the 1D soft wall potential'' in our one-dimensional billiard 
is the closest point that particles approach the turning point. 
In other words, it is the point where the ``penetration'' into the soft 
1D wall is maximal. When particle $1$ has enough momentum ($p_1=-4.0$) to 
``climb''(``penetrate'') the left 1D wall until the ``top'', it reaches 
the turning point on the left (around $q_1\sim -1.1$) with zero momenta 
[see Fig.~\ref{PSS1.0}(a)]. After that, particle $1$ returns to move to 
the right, accelerates ($p_1=+4.0$) and travels until the other extremum 
at $q_1\sim -0.2$. This is a non-periodic {\it chaotic trajectory} with 
a behavior similar to the regular trajectory discussed above. The main 
difference is that the regular trajectory has not enough momentum to 
``climb''(``penetrate'') the 1D wall potential until the ``top''. In this 
example of the chaotic trajectory, the {\it high-energy double collision}
occurs between particle $1$ (high-momentum), particle $2$ and the soft 
1D wall. It is worth to mention that in the triangle description the above 
high- and low-energy double collisions correspond to the high- and low-energy
particle collisions with the corners of the triangle which are located at 
$y=0$.

Figures \ref{PSS1.0}(b)-(c) show a magnification of some regular regions 
from Fig.~\ref{PSS1.0}(a), revealing the existence of the regular islands.
Figure \ref{PSS1.0}(d) shows the PSS for $\gamma=0.8$, which compared to
$\gamma=1.0$ presents a small number of islands. This is in agreement 
with results from Fig.~\ref{dist2}(b), where the number of occurrences 
of the most probable FTLE has not a minimum for  
$\gamma\sim0.8$, and a small number of trapped trajectories is expected.

We conclude this section saying that the softness of the 1D wall decreases 
the mean FTLE. This effect is so strong that it is responsible for the 
small differences observed for $\gamma\lesssim0.4, \gamma\sim0.96$ and 
$\gamma\gtrsim3.0$ between the quasi-hard-limit and the hard wall case 
[see Fig.~(\ref{mean})]. For these values of $\gamma$ the number of 
occurrences of the FTLE from Fig.~(\ref{max}) has a minimum, and a 
larger amount of sticky trajectories is expected. This means that 
the softness of the 1D wall has a stronger influence on space-phases with 
sticky trajectories. In addition we note that the above differences are 
stronger for $\gamma\gtrsim3.0$ [see Fig.~(\ref{mean})]. This can be 
nicely explained using properties of the particle-particle frontal 
collision case. When a frontal collision of the particles occurs, then 
for $\gamma=m_2/m_1>3.0$ the momentum of particle $2$ is large enough
and, after one collision, it continuous to move in the same direction 
as before the collision. Obviously this depends on the particles energy, 
but in average, more frontal collisions are necessary to change the
direction of movement of particle $2$ when $\gamma>3.0$. This property 
increases the amount of double collisions close to the 1D wall, and 
consequently, the effect of the 1D soft wall is more pronnouced when 
$\gamma>3.0$, and the dynamics becomes more regular when compared to 
the 1D hard wall case. 

\subsection{Dependence on the softness parameter  $\sigma$}
\label{sotftness}

\begin{figure}[htb]
 \unitlength 1mm
 \begin{center}
 \includegraphics*[width=7cm,angle=0]{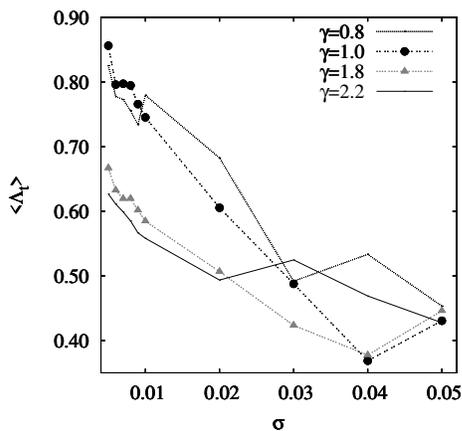}
 \end{center}
\caption{Mean values of the finite-time largest Lyapunov exponent calculated
  over $400$ trajectories up to time $t = 10^4$ as a function of $\sigma$, for
  $\gamma=0.8,1.0,1.8,2.2$.} 
  \label{mean_sg}
  \end{figure}

In this section we analyze the mean FTLEs, $\langle\Lambda_{t}\rangle$, 
and the number of occurrences of the most probable FTLEs as a function 
of the parameter $\sigma$. All simulations were realized over
$400$ initial conditions and for the mass ratios $\gamma=0.80, 1.0, 1.8,
2.2$. Figure \ref{mean_sg} shows  $\langle\Lambda_{t}\rangle$ for $10$
different values of $\sigma$ in the interval
$5 \times 10^{-3}\le\sigma\le5 \times 10^{-2}$. This is exactly the 
transition region between the 1D quasi-hard limit and the 1D soft wall. 
We observe that the main behaviour for all mass ratios is that the mean 
FTLEs decrease as $\sigma$ increases. This means that the degree of 
chaoticity decreases when the softness of the 1D wall increases. 

\begin{figure}[htb]
 \unitlength 1mm
 \begin{center}
 \includegraphics*[width=7cm,angle=0]{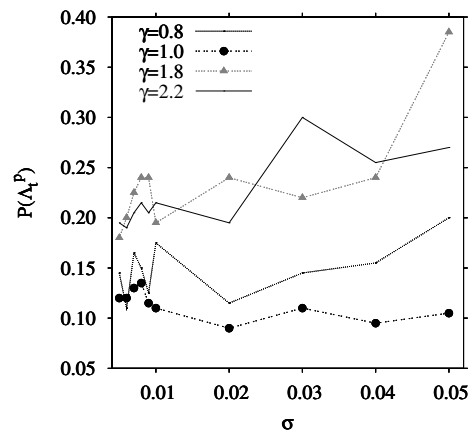}
 \end{center}
\caption{Normalized number of occurrences of the most probable Lyapunov exponent 
  $\Lambda_{t}^{p}$ as a function of $\sigma$.}
  \label{nr_ocur_sg}
  \end{figure}

Figure \ref{nr_ocur_sg} shows the number of occurrences of the most probable 
FTLE as a function of $\sigma$ for the values of $\gamma$ shown in 
Fig.~\ref{mean_sg}.
For all mass ratios, but one (see filled circles $\gamma=1.0$ ),
$P(\Lambda_{t}^{p})$  increases with $\sigma$. This means that the 
dispersion around the most probable FTLE decreases when  $\sigma$ 
increases, i.~e., many initial conditions converge to the same Lyapunov
exponent, and a more ergodic-like motion is expected when compared with 
other mass ratios. This is interesting seeing that for higher values 
of $\sigma$ the degree of chaoticity decreases but the dynamics becomes
more ergodic-like. Compare for example the values of $\Lambda_{t}^{p}$ 
in Fig.~\ref{nr_ocur_sg} for $\sigma=5\times10^{-2}$ and $\gamma=0.8,1.0$.
For $\gamma=0.8$ we have $P(\Lambda_{t}^{p})\sim0.2$, and for $\gamma=1.0$ we 
have $P(\Lambda_{t}^{p})\sim0.1$. So we expect that the dynamics is more 
ergodic-like for $\gamma=0.8$. This really happens and can be confirmed 
visually comparing Figs.~\ref{PSS1.0}(a) and (d). Clearly the 
amount of regular island inside the chaotic sea is smaller for 
$\gamma=0.8$ [Fig.~\ref{PSS1.0}(d)] than for $\gamma=1.0$ 
[Fig.~\ref{PSS1.0}(a)].

\section{Conclusions}
\label{conclusions}

To conclude, we discuss the effect of physical realizable potentials
(soft potentials) on the dynamics of interacting particles inside 
1D billiards. This contribution generalizes previous 
results~\cite{cesar1,cesar2} to the case of 1D soft walls. The 1D soft walls 
are modelled by the error function with 
the softness parameter $\sigma$. In the limit  $\sigma\rightarrow0$ the 1D wall 
force on the particle is given by the $\delta$-function. This allows us to 
study continuously the dynamics of interacting particles in the transition 
from soft to hard walls. The equivalence between the two interacting 
particles in the 1D soft walls model with the motion of one particle inside 
a {\it soft} right triangular billiard is shown. The role of all parameters 
from the model  becomes clear in the right triangular description.
Since the chaotic motion of interacting particles inside 1D billiards is 
generated by double collisions which occurs close to the 1D 
walls~\cite{cesar1}, we expect the influence of the soft 1D walls on the 
particles dynamics to be very strong.  Using the mean FTLE and the number of 
occurrences of the most probable FTLE, we analyze the dynamics of the
interacting particles when the softness parameter changes from the
1D quasi-hard limit ($\sigma=5\times 10^{-3}$) to the 1D soft wall case
($\sigma=5\times 10^{-2}$). We show that the 1D quasi-hard wall limit agrees 
very well with results from the 1D hard wall case, analyzed 
previously~\cite{cesar2}. When the softness parameter increases to 
$\sigma=5\times 10^{-2}$, the mean FTLEs decreases substantiously
(around $50\%$) when compared to the FTLEs from $\sigma=5\times10^{-3}$. 
Although both 1D walls are visually very similar, the dynamics of the 
interacting particles changes considerably. While regular islands and 
trapped trajectories are induced by the {\it low-energy 
double collisions}, the chaotic motion is produced by the {\it high-energy 
double collisions}. Double collisions are characterized by the simultaneous
particle-particle-1D wall collisions. The rise of trapped trajectories
is shown by using the  number of occurrences of the most probable FTLE 
and the corresponding PSS. 

Results from the present paper strongly suggest that the transport of 
interacting particles and heat conduction in physical devices
\cite{kaplan1,kaplan2,weingaertner,barreiro08,wang04,casati20071,casati2007-2,
casati08}, 
which can be described by open billiard models, will substantiously be affected 
by physically realizable wall potentials (soft walls). In such models, 
particles (heat) 
are injected at one open end of the billiard, and the efficiency of the
transport depends how long (among other properties) particles (heat) will 
need to reach the other open end of the billiard. Therefore, trapped 
trajectories induced by the soft walls {\it and } interacting particles, as 
shown in this paper, may substantiously increase the time spend by the 
particles (heat) inside the soft billiard, affecting the transport efficiency.

\begin{acknowledgments}
The authors thank CNPq, CAPES and FINEP, under project CTINFRA-1, for 
financial support.
\end{acknowledgments}


\end{document}